\documentclass{webofc}
\usepackage[varg]{txfonts}   
\usepackage{hyperref}
\usepackage{url}
\hypersetup{colorlinks=true,citecolor=blue,urlcolor=blue,linkcolor=blue}

\RequirePackage{orcidlink}
\usepackage{comment}
\usepackage{multirow}
\usepackage[font=small,labelfont=bf]{caption} 
\newcommand{\code}[1]{\texttt{\small #1}}
\renewcommand{\url}[1]{\href{#1}{#1}}
\renewcommand{\doi}[1]{\href{https://doi.org/#1}{doi:#1}}
\newcommand{\arxiv}[1]{\href{https://arxiv.org/abs/#1}{arxiv:#1}}

\newcommand{\ggttgg}{\mbox{$gg\!\rightarrow\! t\bar{t}gg$}}
\newcommand{\ggttggg}{\mbox{$gg\!\rightarrow\! t\bar{t}ggg$}}
\newcommand{\guxtaptamggux}{\mbox{$g\bar{u}\!\!\rightarrow\!\! \tau^+\!\tau^-\!gg\bar{u}$}}
\newcommand{\ppdyjjj}{\mbox{$pp\!\rightarrow\! \ell^+\!\ell^-\!jjj$}}

\begin{document}

\vspace*{-0.4cm} 
\title{Madgraph on GPUs and vector CPUs: towards production}
\subtitle{The 5-year journey to the first LO release CUDACPP v1.00.00}

\author{
\firstname{Andrea} \lastname{Valassi}\inst{1}
\orcidlink{0000-0001-9322-9565}
\fnsep\thanks
{\email{andrea.valassi@cern.ch}
}
\and
\firstname{Taylor} \lastname{Childers}\inst{2}
\orcidlink{0000-0002-0492-613X}
\and
\firstname{Stephan} \lastname{Hageb\"ock}\inst{1}
\orcidlink{0000-0001-9359-2196}
\and
\firstname{Daniele} \lastname{Massaro}\inst{1}
\orcidlink{0000-0002-1013-3953}
\and\\
\firstname{Olivier} \lastname{Mattelaer}\inst{3}
\orcidlink{0000-0002-7302-7744}
\and
\firstname{Nathan} \lastname{Nichols}\inst{2}
\orcidlink{0000-0002-7386-8893}
\and
\firstname{Filip} \lastname{Optolowicz}\inst{1}
\orcidlink{0009-0001-8723-4365}
\and
\firstname{Stefan} \lastname{Roiser}\inst{1}
\orcidlink{0000-0002-5600-8592}
\and\\
\firstname{Jørgen} \lastname{Teig}\inst{1}
\orcidlink{0009-0002-5618-776X}
\and
\firstname{Zenny} \lastname{Wettersten}\inst{1}
\orcidlink{0000-0001-6099-1799}
}
\institute{
CERN, Geneva, Switzerland
\and
Argonne National Laboratory, USA
\and
Universit\'e Catholique de Louvain, Belgium
}

\abstract{The effort to speed up the Madgraph5\_aMC@NLO generator by exploiting CPU vectorization and GPUs, which started at the beginning of 2020, has delivered the first production release of the code for leading-order (LO) processes in October 2024. 
To achieve this goal, many new features, tests and fixes have been implemented in recent months. 
This process benefitted also 
from the early feedback 
of the CMS experiment.
In this contribution, we report on these activities and on the status of the LO 
software at the time 
of CHEP2024.
\vspace*{-0.3cm}
\vspace*{-0.15cm} 
}

\maketitle

\newcommand{\mgamc}{MG5aMC}

\section{Introduction}
\label{sec:intro}
MadGraph5\_aMC@NLO~\cite{bib:mg5,bib:mg5amc} 
(hereafter, \mgamc) 
is a physics event generator
used in the data processing workﬂows 
of many High Energy Physics (HEP) experiments, 
notably those at CERN’s Large Hadron Collider (LHC).
The computational cost of event generation
is a non-negligible fraction of the total 
computing costs of the LHC experiments
and is predicted to further increase 
as the high-luminosity LHC programme (HL-LHC) gets underway~\cite{bib:csbs}.
The \mgamc\ software 
has been developed over more than two decades
and initially targeted CPUs only, mostly
using sequential processing paradigms.
Over time, new computing architectures
designed for parallel processing,
such as multi-core CPUs with wide vector registers
and graphical processing units (GPUs)
have become widely available in the 
Worldwide LHC Computing Grid (WLCG) infrastructure.
This represents a challenge,
because 
porting legacy software 
to efficiently support these new architectures
require their radical rethink and redesign,
and failure to do so may lead to large
under-utilization of the computing power 
of existing 
hardware resources;
but it also poses an opportunity,
as the efficient exploitation of these resources
using better software can significantly reduce
the computing costs of the LHC experiments.
This was precisely the motivation
behind the development~effort 
described in this paper,
which started in February 2020 
with the initial goal of porting \mgamc\ to GPUs,
but was very soon extended
to also optimize the code for vector CPUs,
using the same data parallel approach
and the same code base
for both architectures.
After almost five years of development,
the first production release of the code
was delivered in October 2024,
just before the CHEP2024 conference
where we have presented it 
for the first time~\cite{bib:chep2024slides}.

\mgamc\ is a code generating framework,
largely written in Python,
which allows users to 
generate and run physics code 
to perform 
Monte Carlo (MC) event generation 
for any physics process of their choice.
By default, physics code in \mgamc\ is
generated in Fortran
and can only run on CPUs,
with no support for vectorization:
this is the production backend 
that has been used so far by the LHC experiments 
in their event generation campaigns.
The main deliverable 
of the work described in this paper
is a new code-generating plugin 
that we 
named ``CUDACPP'', 
because the generated physics code
uses C++ instead of Fortran
to execute the most computationally intensive
sections of the code on vector CPUs,
and also includes CUDA extensions 
to run the code on NVidia GPUs.
Our contribution
includes not only the first release
v1.00.00 of the CUDACPP plugin,
but also the addition
of many changes that were needed 
in the overall framework
into a new production release
\mgamc\ v3.6.0.

In this paper,
we focus on the description
of the additional work that was necessary
to achieve the release,
with respect to what 
has been presented 
at previous conferences~\cite{bib:vchep2021,bib:ichep2022,bib:acat2022,bib:chep2023}.
This 
includes new features
such as the support for 
Beyond-Standard-Model (BSM) processes,
the support for AMD GPUs 
via HIP language extensions
to our previous CUDA/C++ codebase, 
a mechanism to 
distribute our software as a new plugin
of the 
\mgamc\ framework,
and especially the extensive
testing, debugging and optimization
of the full \mgamc\ workflow
and of these new functionalities.
Work is also in progress 
to support Intel GPUs
via SYCL language extensions,
integrating into the CUDACPP plugin
the prototypes we 
presented at previous conferences~\cite{bib:ichep2022,bib:acat2022,bib:chep2023},
but we will not cover any of this work
in this paper.
Our work on 
the first CUDACCP release
has also benefitted from the early feedback
of our colleagues from 
the CMS experiment,
whose extensive tests
of the functionality and performance speedup
achievable using our software
have also been presented 
to this conference~\cite{bib:jin}.
Currently, CUDACPP only supports
Leading-Order (LO) physics processes:
The techniques and software developed
for LO 
are largely reusable also 
for Next-to-Leading-Order (NLO) processes.
Extending CUDACPP to also support NLO
is one of the main further directions
of ongoing developments and
is described in a separate 
contribution to this conference~\cite{bib:zenny}.

This is the outline of this paper:
in Sec.~\ref{sec:arch} we describe
the architecture and
the evolving focus 
of developments over time;
in Sec.~\ref{sec:perf} 
we give performance results
and an outlook. 

\newcommand{\tabcppggttggg}[1]{
\begin{table}[#1]
\vspace*{-0.3cm}
\begin{center}{
\small
\hspace*{-4mm}
\setlength\tabcolsep{5pt} 
\begin{tabular}{|l|c|c|c|c|}
\cline{3-5}
\multicolumn{2}{c|}{}& 
\multicolumn{3}{c|}{madevent} \\ 
\hline
\multirow{2}{*}{\ggttggg} & MEs &
$t_\mathrm{TOT} = t_\mathrm{Mad} + t_\mathrm{MEs}$ &
$N_\mathrm{events} / t_\mathrm{TOT}$ &
{$N_\mathrm{events} / t_\mathrm{MEs}$}\\
& precision &
[sec] &
[events/sec] &
[MEs/sec] \\
\hline
Fortran(scalar) & double &
854.1 = 2.8 + 851.3 &
9.59E1 (=1.0) &
9.62E2 (=1.0) \\
\hline
C++/none(scalar) & double&
970.8 = 3.0 + 967.9 &
8.44E1 (x0.9) &
8.46E1 (x0.9) \\
C++/sse4(128-bit) & double &
506.8 = 2.9 + 503.9 &
1.62E2 (x1.7) &
1.63E2 (x1.7) \\
C++/avx2(256-bit) & double &
235.3 = 2.8 + 232.5 &
3.48E2 (x3.6) &
3.52E2 (x3.7) \\
C++/512y(256-bit) & double &
209.6 = 2.8 + 206.8 &
3.91E2 (x4.1) &
3.96E2 (x4.1) \\
C++/512z(512-bit) & double &
116.8 = 2.8 + 114.0 &
7.01E2 (x7.3) &
7.19E2 (x7.5) \\
\hline
C++/none(scalar) & mixed &
983.6 = 3.0 + 980.6 &
8.33E1 (x0.9) &
8.35E1 (x0.9) \\
C++/sse4(128-bit) & mixed &
491.5 = 2.9 + 488.7 &
1.67E2 (x1.7) &
1.68E2 (x1.7) \\
C++/avx2(256-bit) & mixed &
227.8 = 2.8 + 199.2 &
3.60E2 (x3.8) &
3.64E2 (x3.8) \\
C++/512y(256-bit) & mixed &
202.0 = 2.8 + 199.2 &
4.05E2 (x4.2) &
4.11E2 (x4.3) \\
C++/512z(512-bit) & mixed &
116.6 = 2.8 + 113.8 &
7.03E2 (x7.3) &
7.20E2 (x7.5) \\
\hline
C++/none(scalar) & float &
943.5 = 3.0 + 940.6 &
8.68E1 (x0.9) &
8.71E1 (x0.9) \\
C++/sse4(128-bit) & float&
229.8 = 2.8 + 227.0 &
3.56E2 (x3.7) &
3.61E2 (x3.7) \\
C++/avx2(256-bit) & float&
118.9 = 2.8 + 116.0 &
6.89E2 (x7.2) &
7.06E2 (x7.3) \\
C++/512y(256-bit) & float&
106.7 = 2.8 + 103.9 &
7.68E2 (x8.0) &
7.89E2 (x8.2) \\
C++/512z(512-bit) & float&
\hphantom{0}60.6 = 2.8 + \hphantom{0}57.8 &
1.35E3 (x14.1) &
1.42E3 (x14.7) \\
\hline
\end{tabular}
}\end{center}
\vspace*{-6mm}
\caption{
Processing times 
(ME, non-ME, total)
and throughputs 
(total, ME)
for 
81920 \ggttggg\ weighted events.
CERN itgold91
with Intel Gold 6326 CPUs,
using gcc11.4 builds.
In this and the following tables,
madevent results refer to
the execution of one application
on a single CPU core.
}
\label{tab:gold}
\vspace*{-0.5cm}
\end{table}
}

\newcommand{\tabcudaggttggg}[1]{
\begin{table}[#1]
\begin{center}{
\small
\hspace*{-2mm}
\setlength\tabcolsep{5pt} 
\begin{tabular}{|l|c|c|c|c|c|}
\cline{3-6}
\multicolumn{2}{c|}{}& 
\multicolumn{3}{c|}{madevent}& 
\multicolumn{1}{c|}{standalone}\\
\hline
\multicolumn{2}{|c|}{CUDA grid size}& 
\multicolumn{3}{c|}{8192}& 
\multicolumn{1}{c|}{16384}\\
\cline{1-6}
\multirow{2}{*}{\ggttggg} &
MEs &
$t_\mathrm{TOT} = t_\mathrm{Mad} + t_\mathrm{MEs}$ &
$N_\mathrm{events} / t_\mathrm{TOT}$ &
\multicolumn{2}{c|}{$N_\mathrm{events} / t_\mathrm{MEs}$}\\
& precision &
[sec] &
[events/sec] &
\multicolumn{2}{c|}{[MEs/sec]}\\
\hline
Fortran & double &
\hphantom{0}998.1 = 4.4 + \hphantom{0}993.7 &
8.21E1 (=1.0) &
8.24E1 (=1.0) &
--- \\
\hline
CUDA/GPU & double &
\hphantom{00}16.8 = 5.9 + \hphantom{00}10.9 &
4.88E3 (x60) &
7.54E3 (x92) &
9.54E3 (x115) \\ 
\hline
CUDA/GPU & mixed &
\hphantom{00}14.3 = 5.7 + \hphantom{000}8.6 &
5.72E3 (x70) &
9.49E3 (x115) &
1.16E4 (x141) \\
\hline
CUDA/GPU & float &
\hphantom{00}10.7 = 5.4 + \hphantom{000}5.3 &
7.65E3 (x94) &
1.53E4 (x187) &
2.16E4 (x264) \\
\hline
\end{tabular}
}\end{center}
\vspace*{-6mm}
\caption{
Processing times and throughputs 
for 
81920 \ggttggg\ weighted events.
Single core of CERN itscrd90
with Intel Silver 4216 CPUs
and NVidia V100 GPU,
using gcc11.3
and nvcc12.0 builds.
ME throughputs using 
a larger CUDA grid size
in the standalone application 
are also shown
for comparison.
}
\label{tab:rd90}
\vspace*{-0.8cm}
\end{table}
}

\newcommand{\tabcppdyj}[1]{
\begin{table}[#1]
\vspace*{-0.3cm}
\begin{center}{
\small
\hspace*{-4mm}
\setlength\tabcolsep{5pt} 
\begin{tabular}{|l|c|c|c|c|}
\cline{3-5}
\multicolumn{2}{c|}{}& 
\multicolumn{3}{c|}{madevent}\\
\hline
\guxtaptamggux & MEs &
$t_\mathrm{TOT} = t_\mathrm{Mad} + t_\mathrm{MEs}$ &
$N_\mathrm{events} / t_\mathrm{TOT}$ &
{$N_\mathrm{events} / t_\mathrm{MEs}$}\\
(81920 weighted events)
& precision &
[sec] &
[events/sec] &
[MEs/sec] \\
\hline
Fortran(scalar) & double &
52.2 = 17.0 + 35.2 &
1.57E3 (=1.0) &
2.32E3 (=1.0) \\
\hline
C++/none(scalar) & mixed&
50.9 = 16.9 + 33.9 &
1.61E3 (x1.0) &
2.41E3 (x1.0) \\
C++/sse4(128-bit) & mixed &
33.9 = 16.9 + 17.0 &
2.41E3 (x1.5) &
4.82E3 (x2.1) \\
C++/avx2(256-bit) & mixed &
24.8 = 17.2 + \hphantom{0}7.6 &
3.31E3 (x2.1) &
1.08E4 (x4.7) \\
C++/512y(256-bit) & mixed &
24.1 = 17.1 + \hphantom{0}7.0 &
3.40E3 (x2.2) &
1.18E4 (x5.0) \\
C++/512z(512-bit) & mixed &
26.5 = 17.0 + \hphantom{0}9.6 &
3.09E3 (x2.0) &
8.57E3 (x3.7) \\
CUDA/GPU & mixed &
17.7 = 17.4 + \hphantom{0}0.3 &
4.64E3 (x3.0) &
3.23E5 (x138) \\
\hline
\end{tabular}
}\end{center}
\vspace*{-5mm}
\caption{
Processing times and throughputs 
for 
81920 \guxtaptamggux\ weighted 
events. \!Single core of CERN itscrd90
with Intel Silver 4216 CPUs
and NVidia V100 GPU,
using gcc11.3
and nvcc12.0 builds.
}
\label{tab:dy3j}
\end{table}
}

\newcommand{\tabcpplumi}[1]{
\begin{table}[#1]
\vspace*{-0.3cm}
\begin{center}{
\small
\hspace*{-4mm}
\setlength\tabcolsep{5pt} 
\begin{tabular}{|l|c|c|c|c|}
\cline{3-5}
\multicolumn{2}{c|}{}& 
\multicolumn{3}{c|}{madevent}\\
\hline
\multirow{2}{*}{\ggttgg} & MEs &
$t_\mathrm{TOT} = t_\mathrm{Mad} + t_\mathrm{MEs}$ &
$N_\mathrm{events} / t_\mathrm{TOT}$ &
{$N_\mathrm{events} / t_\mathrm{MEs}$}\\
& precision &
[sec] &
[events/sec] &
[MEs/sec] \\
\hline
Fortran(scalar) & double &
26.6 = 1.4 + 25.2 &
3.09E3 (=1.0) &
3.25E3 (=1.0) \\
\hline
C++/none(scalar) & mixed&
33.2 = 1.4 + 31.8 &
2.47E3 (x0.8) &
2.57E3 (x0.8) \\
C++/sse4(128-bit) & mixed &
16.7 = 1.4 + 15.3 &
4.91E3 (x1.6) &
5.36E3 (x1.6) \\
C++/avx2(256-bit) & mixed &
\hphantom{0}8.3 = 1.4 + \hphantom{0}6.9 &
9.93E3 (x3.2) &
1.20E4 (x3.7) \\
HIP/GPU & mixed &
\hphantom{0}2.9 = 1.8 + \hphantom{0}1.1 &
2.88E4 (x9.3) &
7.69E4 (x24) \\
\hline
\end{tabular}
}\end{center}
\vspace*{-5mm}
\caption{
Processing times and throughputs 
for 
81920 \ggttgg\ weighted events.
Single core of 
LUMI HPC
with AMD EPYC 7A53 CPUs
and AMD Instinct MI200 GPUs,
using gcc13.2
and hipcc6.0 builds.
}
\label{tab:lumi}
\vspace*{-0.6cm}
\end{table}
}

\vspace*{-0.12cm}
\section{Overview of architecture and developments}
\label{sec:arch}
\vspace*{-0.08cm}

\begin{figure}[t]
\vspace*{-0.30cm}
\centering
\hspace*{-0.02\textwidth}
\includegraphics[width=1.03\textwidth,clip]{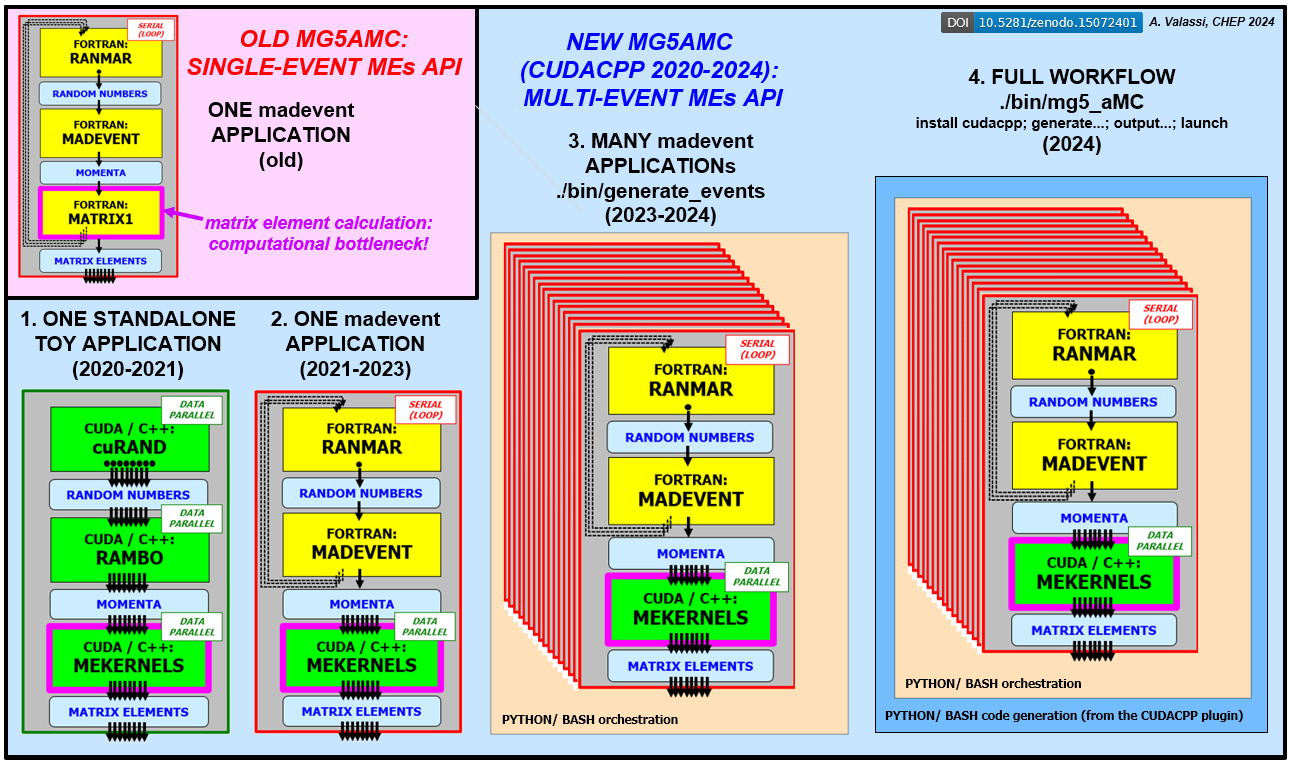}\\
\caption{
Schematic representation
of the architectural evolution
of \mgamc.
The main difference between
the old Fortran-only version
(top left, pink background)
and those based on CUDACPP
(light blue background)
is that the former uses 
a sequential single-event API
for the calculation of matrix elements,
while the latter uses 
a data-parallel multi-event API.
Additional details on the
evolution of the work
on the 
CUDACPP plugin
between 2020 and 2024
are provided in the text.
These plots, presented at 
CHEP2024~\cite{bib:chep2024slides},
are derived from those
presented in a 2020 talk
to the HSF generator WG~\cite{bib:hsf2020}.
}
\vspace*{-0.8cm}
\label{fig:anatomy}
\end{figure}

From a computational point of view,
\mgamc\ is a 
complex framework 
with many different software layers.
What follows is only a high-level summary;
a more complete description can be found
in the 
\mgamc\ paper~\cite{bib:mg5amc}.
Essentially,
a user who wants 
to compute a cross section 
or generate some events 
mainly needs to specify 
the desired physics process,
the required precision or sample size
and the desired backend,
via an interactive prompt 
or a batch script.
The \mgamc\ framework
(largely written in Python) 
then takes care of everything else:
it determines the subprocesses
that must be computed
to handle the desired physics process;
it generates physics code
for the required backend
(by default, using Fortran for CPUs)
and builds it into
one ``madevent'' executable
for each subprocess;
it optimizes 
configuration files
via test runs
and optionally creates a ``gridpack''
for Grid distribution;
it executes one or more
copies of all relevant
madevent executables;
finally, it combines
the results of the different
program executions to provide 
the overall output cross section
and/or an output LHE file
aggregating all MC events
that have been generated.
Since the first software commits
for our project in February 2020,
achieving the delivery 
of the first production release
took almost five years
because we had to 
adapt and test 
all of these layers.
We did this incrementally
in successive steps,
as described
in the following.
This is schematically 
represented 
in Fig.~\ref{fig:anatomy}.

Internally,
each madevent executable
has itself a complex structure.
A similar 
data flow
is~found~\cite{bib:hsf2020}
in any matrix element generator (MEG),
not just in \mgamc.
Schematically,
all calculations imply the 
generation of 
events using MC techniques,
in three steps:
first, for each event,
some random numbers are drawn
using a pseudo-random number generator;
second, these random numbers are converted 
into particle momenta
using a phase space sampling algorithm;
third, from the particle momenta,
the ``matrix element'' 
(ME), essentially the probability of the event
for the given physics process, 
is computed.
In the version of \mgamc\ that 
has been used prior to our work,
and in most of the existing MEGs,
this is carried out
using serial processing,
i.e. a global event loop
where these three steps
are performed in sequence
for one event at a time.
This is described in the top-left diagram
in Fig.~\ref{fig:anatomy}.

Alternatively, however,
each of the three steps above
can be performed in parallel 
for many events at a time.
In particular,
as it was pointed out within 
the HSF generator working group~\cite{bib:csbs,bib:hsf2020,bib:lhcc2020},
event-level data parallelism
is an approach
that may allow a very efficient 
use of GPUs or vector CPUs in any MEG.
This is because,
to a large extent, 
the same mathematical functions
need to be numerically computed
for different events
in each of these three steps.
The near absence of stochastic branching,
which ensures near perfect 
lockstep processing 
of many events at a time,
is in fact essential to efficiently
exploiting SIMD 
(Single Instruction Multiple Data) on
vector CPUs via vectorized code,
and it also significantly enhances
the efficiency of SIMT
(Single Instruction Multiple Threads)
on GPUs by reducing thread divergence.
This is especially true and important
for the third step 
of a MEG, 
the computation of matrix elements,
because this is by far the most 
computationally intensive bottleneck
in these calculations 
for non-trivial physics processes,
and accelerating 
the ME calculation alone
allows significant speedups
of the whole workflow.
This is precisely the goal
that we have pursued and achieved 
in our work 
on the CUDACPP plugin of \mgamc.

The first step of our work
in 2020-2021~\cite{bib:vchep2021}
initially focused on 
a standalone toy application,
fully coded in C++ and CUDA,
where also random number generation
and phase space sampling,
not only ME calculations,
follow a data-parallel approach.
This is 
represented as diagram 1 in 
the bottom-left corner of 
Fig.~\ref{fig:anatomy}.
Its goal 
is to provide a 
fast test framework where 
the ME calculation kernels
(in CUDA for GPUs or vectorized C++ for CPUs)
can be debugged and optimized.
The standalone application
is still used in this way today
by the~development team.
Another 
reason 
to maintain it
is that it is useful to create 
a CUDACPP ME library,
for reweighting~\cite{bib:reweight}
or for its integration
into frameworks other than \mgamc. 
Code generation and
AOSOA data structures for vectorization
were completed in this phase.

In a second step,
around 2021-2023~\cite{bib:ichep2022,bib:acat2022,bib:chep2023},
we moved to
the integration of 
the CUDA/C++ ME kernels in 
the Fortran madevent application
(diagram 2 in Fig.~\ref{fig:anatomy}).
An essential 
point
was moving
madevent 
from a single-event
to a multi-event API for MEs,
ensuring that computational kernels
are 
reentrant stateless functions
with well-defined input and output data.
Doing this will also be 
critical in
the GPU port 
of other MEGs or 
of NLO processes in \mgamc.
Also important
was the development 
of a new large 
suite of functional and performance tests
around a single madevent application.
The functional tests 
check
that, within numerical precision,
the same 
LHE event file and cross section
are obtained
from the same random number seeds,
independently of whether 
MEs are computed in Fortran, CUDA or C++,
and in single, double, or "mixed" 
precision.
These tests 
are
routinely
executed for many 
physics processes and, in 2024, 
were also
integrated 
in the project's github CI.
Together with complementary tests
focusing on possible regressions
in new features in madevent,
they 
have been invaluable for 
identifying 
and fixing 
different types 
of bugs
up until the release.
The performance tests, similarly,
have made it 
possible to compare
the processing times 
of one
madevent application
using different backends and 
precisions for the ME calculation.
The performance numbers that we give in 
Tables~\ref{tab:gold}, \ref{tab:rd90},
\ref{tab:lumi} and~\ref{tab:dy3j},
like many of those presented 
at previous conferences,
have been produced with this infrastructure.
These tests are particularly useful
because they separately measure
the time taken 
by the ME calculation
(which significantly decreases 
for data-parallel backends)
and the time taken
by other non-ME components of the
madevent executable
(which is roughly the same
for all ME backends,
as these are serial calculations).

The last steps of our work since 2023,
which we have not presented 
at previous conferences,
has focused on the 
integration, testing and optimization
of the full \mgamc\ workflow.
Initially
(diagram 3 in Fig.~\ref{fig:anatomy}),
we tested the launching
of many 
madevent executables
from a gridpack that we had 
prepared by manually generating
and building the relevant physics code.
One example is given
in Fig.~\ref{fig:dy3j}.
This is useful to understand 
how CUDACPP-generated code behaves
in the overall workflow.
One delicate issue, 
in particular,
is that the event-level parallelism
in this first release of CUDACPP
requires a large number of events
(typically, 8k or 16k)
to be processed 
at the same time
in a CUDA grid
for GPUs to be efficient,
while the \mgamc\ framework
was initially designed 
to launch many madevent executables
each processing O(1000) events.
Because of the complex strategy
followed in \mgamc\ to cover
all different ``channels''
of phase space integration
(where one channel roughly 
corresponds to the peaking structure
of a single Feynman diagram),
launching too many events
in a single madevent executable
may lead to physics biases.
One approach that we have followed
to mitigate this issue has been to modify
CUDACPP to allow the processing 
of events from different channels
in a single ME kernel
(while preserving lockstep 
by processing only events 
from the same channel
in any given GPU warp or CPU SIMD lane).
A second, complementary, approach 
consists instead in going beyond
pure event-level parallelism
and processing in parallel
not only different events,
but also different helicity combinations
of the particles from each event.
This latter 
work is recent and 
not yet committed upstream:
preliminary 
results~\cite{bib:ks2024} are however promising,
as they indicate
that this approach may reduce 
by one to two orders
of magnitude the number of events
that must be processed in parallel
by each madevent executable,
especially if
different CUDA streams are used
for different helicities.

\tabcppggttggg{t}
\tabcudaggttggg{t}

The very last step of our work
before the release in 2024
(diagram 4 in Fig.~\ref{fig:anatomy})
has been to ensure
that the CUDACPP code-generating plugin
is seamlessly integrated 
in the \mgamc\ end-user experience.
One practical problem 
to address was 
that our work
on the 
\mgamc\ framework
and on CUDACPP has
been carried out
using two separate code
repositories, currently 
mg5amcnlo~\cite{bib:mg5amc-github}
and madgraph4gpu~\cite{bib:mg4gpu-github}.
However, many 
developments
require
changes on both sides,
and a mechanism is needed
to ensure that any development branch
in one repository is used 
against a compatible branch 
in the other repository.
For the purpose of 
our internal development work,
since 2023 we have used
github submodules for 
this synchronization.
In particular,
every branch of
madgraph4gpu includes 
a specific commit of
mg5amcnlo as a submodule.
For the purpose 
of the end-user experience,
we considered initially
a similar option
where mg5amcnlo 
includes the CUDACPP plugin from madgraph4gpu
as a submodule,
as well as a more monolithic second option
where the CUDACPP plugin is 
moved into
a subdirectory of mg5amcnlo.
We finally settled
for a third option,
which is 
more consistent
with how other 
plugins are handled within \mgamc:
we kept the two repositories 
as they are
(i.e. madgraph4gpu still includes 
mg5amcnlo as a submodule),
but we added to madgraph4gpu
a mechanism to prepare a tarball
of the CUDACPP plugin
whenever a release tag is created,
and we configured mg5amcnlo
to 
download and use
the CUDACPP tarball if required.
Users only need to add
\code{install cudacpp} to their
\mgamc\ batch scripts,
and choose the CUDA or C++ backend,
to be able to use our work.
In the future,
it is 
likely that we will clean up and rename
the madgraph4gpu repository,
while maintaining this tarball mechanism
and the rich history 
of issues and pulls requests
documenting the work so far.

Before moving to performance results 
in the next section,
a few additional details
on two important areas of work in 2023-2024
should 
be mentioned.
First: a large number of
Beyond-Standard-Model (BSM) physics processes,
notably from SUSY, SMEFT and HEFT models,
have been extensively tested and debugged.
This is important because these are processes
that the LHC experiments typically 
simulate at LO rather than at NLO,
and where they explore very large 
numbers of BSM parameter configurations:
therefore, completing support for them
in the current LO CUDACPP plugin
may provide significant cost reductions.
However, and more importantly, 
this has also been essential
to identify, debug and fix
various issues 
affecting SM processes,
which would have otherwise gone undetected.
In particular, many subtle bugs
involving Floating Point Exceptions
in SIMD code have been addressed. 
Second: support for AMD GPUs 
has been added to the CUDACPP plugin
via HIP language extensions
to our previous CUDA/C++ codebase.
We used an approach
based on \#ifdef directives,
similar to what we do to
distinguish between CPU and GPU sections
in CUDACPP-generated physics code.
This was relatively easy because
the CUDA and HIP APIs are very similar:
for instance,
we defined
a generic function \code{gpuMalloc}
as either
\code{cudaMalloc} or \code{hipMalloc}.
Adding support for Intel GPUs
is instead more complex because
the SYCL API uses different concepts.
Support for AMD GPUs (and AMD CPUs) has
been 
tested
using resources
at the LUMI HPC centre.
Performance results 
for some of these tests
are shown in Table~\ref{tab:lumi}. 

\tabcpplumi{t}

\section{Performance results and outlook}
\label{sec:perf}

The performance speedups achievable
in CUDACPP
thanks to ME acceleration 
on GPUs and 
vector CPUs
were already
discussed in 
previous 
papers~\cite{bib:vchep2021,bib:ichep2022,bib:acat2022,bib:chep2023}.
There were only few~performance improvements
in the first CUDACPP release,
and the same general comments 
still apply.
The latest results 
for our $\ggttggg$ standard candle,
for instance, are given
in Table~\ref{tab:gold}
for an Intel Gold CPU 
and in Table~\ref{tab:rd90} 
for an NVidia V100 GPU.
On the CPU,
we see ME speedups
close to factors 8 and 16
in double and single precision,
which are the theoretical limits 
for this hardware
supporting AVX512 SIMD 
with two FMA 
units~\cite{bib:ichep2022}. 
On the GPU,
higher ME speedups 
by factors 90 and 180 
are seen for doubles and floats
(with 8k events per CUDA grid).
For a madevent executable,
speedups are almost unchanged on the CPU
and stay as high as 60 and 90 on the GPU.
As 
the serial non-ME part 
only takes $\sim$1/200 
of the time, 
the maximum speedup 
allowed by Amdahl's law~\cite{bib:vchep2021} 
is $\sim$200:
for a complex~physics process 
like this one,
ME calculations
remain the bottleneck
even after accelerating them.

\tabcppdyj{t}

\begin{figure}[t]
\vspace*{-0.3cm}
\centering
\hspace*{-0.015\textwidth}
\mbox{
\includegraphics[width=0.495\textwidth,clip,clip]{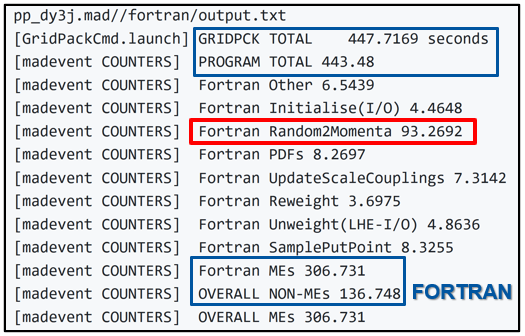}
\includegraphics[width=0.495\textwidth,clip,clip]{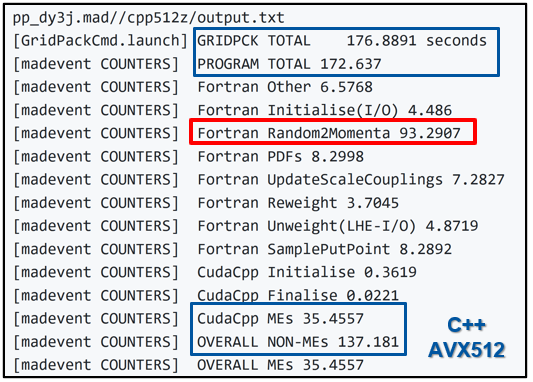}}
\caption{
Breakdown of the ME 
and various non-ME contributions
to the overall runtime
of a DY+3jets (\ppdyjjj)
gridpack,
using Fortran MEs (left)
or CUDACPP "512z" C++ MEs (right).
Gridpack launched on
CERN itgold91
with Intel Gold 6326 CPUs,
using gcc11.4 builds.
The numbers refer
to the generation 
of 100 unweighted events:
this involved the 
execution of 108 madevent applications,
each processing 16384 ME calculations,
for an overall total 
of 1.8M ME calculations 
on weighted events.
}
\label{fig:dy3j}
\vspace*{-0.8cm}
\end{figure}

As we already knew,
numerical precision is
thus extremely important.
In 2023, we 
performed a detailed 
analysis~\cite{bib:filip}
using the CADNA~\cite{bib:cadna} tool,
which made it possible
to identify the sections
of our code where numerical precision
risks to be lost by rounding
and cancellations.
Not surprisingly, these are all
in the generated 
physics code
computing Feynman diagrams.
In the first CUDACPP release,
we thus decided to use as default precision
our ``mixed'' mode,
where Feynman diagrams are computed
in double precision
while the color matrix quadratic form
is computed in single precision.
As in the past,
this is slightly faster 
than double precision on GPUs,
but still essentially equivalent
on vector CPUs:
this may be an intrinsic
limitation, or it may
indicate 
that our SIMD mixed mode
could be improved.

A very important aspect
of our work in 2024
has been our collaboration 
with the CMS experiment,
who have tested our code
and provided extremely valuable feedback
even before its official release.
In particular,
CMS have performed 
tests on their side,
described in their own 
CHEP2024 contribution~\cite{bib:jin},
and have reported many 
observations and
functional and performance issues
to us,
so that we could further investigate 
and often fix them on our side.

An active
area of work 
has been
the study of the performance speedups
achievable for Drell-Yan processes,
which are
relevant to CMS already at LO.
Detailed profiling has shown 
for instance that for DY+3jets (\ppdyjjj)
the ME calculation 
in the original Fortran code
only accounts for around 2/3
of the total time spent.
Since the non-ME components
taking up the remaining 1/3
are currently not parallelized,
Amdahl's law dictates that 
the maximum achievable speedup
for the overall workflow 
is only a factor 3.
At the level of one individual
madevent executable, 
this was tested 
for one particular subprocess
of DY+3jets, namely \guxtaptamggux:
the results
in Table~\ref{tab:dy3j}
indicate that the overall speedup
achieved is only 3.0 for CUDA
and 2.1 for AVX2,
even if the ME speedups
are 140 and 4.7, respectively.

In order to better understand this 
for a
full workflow with 
many madevent executions,
an enhanced profiling infrastructure was 
developed~\cite{bib:prof2024}:
this makes it possible to dump
profiling information 
from each madevent execution,
and 
aggregate it at the end.
This work also 
involved a 
more fine-grained
profiling of each madevent executable
via code instrumentation,
by roughly decomposing
the non-ME time into subcomponents,
such as phase space sampling,
PDF evaluations, unweighting and so on.
With respect 
to flamegraph sampling profiling,
which we also routinely use,
this complementary approach
makes it possible to 
collect data programmatically,
enabling easier comparisons between backends.
This infrastructure is still preliminary
and not yet committed to our code base.
One example is shown in Fig.~\ref{fig:dy3j},
where DY+3jets aggregated time profiles
are compared for the Fortran 
and C++/AVX2 ME backends.
In this specific case,
after accelerating MEs with CUDACPP,
the new computational bottleneck
is phase space sampling.
This is
highly process-dependent,
as there may be processes 
where PDF evaluations are the
new bottleneck instead,
and for complex processes 
MEs may remain the bottleneck.
In any case,
as our work on 
accelerating MEs
has now reached production quality,
identifying
and speeding up
the next bottlenecks
will certainly be 
one of our priorities
for future developments
on LO calculations.
Initial studies in this direction
have already started for
phase space sampling~\cite{bib:prof2024},
showing that data parallelization
should be possible but difficult,
and that there may also be 
some lower-hanging fruits.
Work is also well underway~\cite{bib:pdfs}
towards profiling and speeding up 
PDF evaluations in \mgamc.

\vspace*{-0.2cm}
\section*{Acknowledgements}
\vspace*{-0.2cm}
We kindly 
acknowledge the use 
of LUMI HPC resources 
under project 465001114 
(EHPC-BEN-2024B04-053,
CERN/HEPiX Benchmarking) 
for our results on
AMD CPUs/GPUs.
It is a pleasure to 
thank our CMS colleagues~\cite{bib:jin}
for our 
mutually beneficial
collaboration.

\end{document}